# A CYCLIC UNIVERSE WITHOUT MISSING MASS: IMPLICATIONS OF R → α'/R

WAYNE R. LUNDBERG
*U. S. Air Force, 1801 Tenth St*
*Wright-Patterson AFB, OH 45433-7626*

## ABSTRACT

Results from string theory conclude that a spatial dimension R is equivalent to $\alpha'/R$. If R is considered as a four-space dimension, several interesting results emerge. In this paradigm, $\alpha'/R$ exists in a time-reversed, "anti-parallel" universe. Cosmological distances in our real-time universe are equivalent to the compact dimensions at the instant of a Big Bang in a time-inverted universe. This model agrees with standard black hole theory in that particles become frozen in time when they reach the 'singularity' (of dimension $\alpha'$). A particle which enters a black hole in real time exits from the Big Bang or "white hole" in the time-inverted universe, *leaving behind a trace of annihilation radiation*. A diagram of this phenomenon is constructed, consistent with existing knowledge of the early universe. The primordeal black holes predicted by Hawking are required. A cyclic universe is described by $M^2_{p,n} + T^2_n \approx \Omega_m + \Omega_\Lambda = 1$. The "missing mass" proscribed by standard closed-universe theories is not required. This result is useful if experimental evidence in the search for dark matter continues to leave a significant "missing mass" - *and particularly in light of* **more recent observations indicating that, in fact, inflation continues at a much reduced rate.**

## Introduction

The success of the Big Bang / cold dark matter model has led to numerous hypotheses regarding the physical nature of, and proposed experiments to find, the dark matter required for a closed (but not cyclic) universe. This paper is written to propose that the universe is indeed cyclic ( ∴ 'closed' but flat), and that it's cyclic behavior is a direct consequence of the underlying structure of space-time. Recent theoretical *and experimental* results have emerged which are crucial to understanding the structure of space-time.

String theory has had marked success in establishing a mathematical framework which will allow the unification of particle physics with general relativity. It has yet to be formulated at low enough energies to reproduce exactly the array of particles and interactions known to physicists. This deficiency is of no immediate concern for the discussion of cosmology. The implications of R↔α'/R are quite obviously important and must ultimately modify the standard cosmological model if one comprehensive model is to result.

Edward Witten discusses the structure of space-time as derived from the particle theorist's view[1], particularly that string theory requires a minimum length of $\sqrt{\alpha'} \cong 10^{-32}$ cm. A strikingly similar dimensional arguement arises in Hawking's[2] discussion of the thermal radiation from black holes: the unit of area required is $10^{-66}$ cm$^2$. Although these results may be regarded as coincidental, one thing is certain - that the notion of a true mathematical singularity at the center of a black hole must be re-examined in light of the results from string theory. Penrose[2] agrees in saying "A true theory of quantum gravity should replace our present concept of spacetime at a singularity. It should give us a clear-cut way of talking about what we call a singularity." We shall return to this concept after developing a conceptual framework for a spacetime duality.

## Spacetime Duality

The greatest implication of this peculiar spatial duality comes when one considers how spacetime (as characterized by R) and it's dual (characterized by $\alpha'/R$) behave. It is reasonable to link time with radius since all cosmological models consistent with observations include an ever-expanding universe. Thus the dual of our universe under $R \leftrightarrow \alpha'/R$ is also a time-inverted universe which is continually compacting in our usual reference frame – *quite the opposite of the proposition of a holographic universe*. The extraordinarily small, but non-zero, dimensions required are dismaying until one considers that every particle in our universe contains a tiny reflection of a time-inverted space. High energy particles propagating backward in time simply emerge from normal particles in this model. Linkage at such a fundamental level of particle physics indicates that anti-matter comprises the "stitches" which link our space-time to its dual.

With this dual space-time in mind, it is also much easier to grasp how the matter involved in the Big Bang achieved its initial state: the initial state of our universe is simply the final state of a dual space-time universe. This is quite consistent with the inflationary universe[3] which originates from a state with ultra-compact dimensions equivalent to $<\alpha'$. The final state of our universe occurs when all the black holes in our universe accrete all the matter available to them even if they are spatially vastly separated. The final phase of this model would take an extraordinarily long time, during most of which the universe has very little matter not compacted in black holes. The appropriate scale for universal time is then $\ln T \leftrightarrow \ln(\alpha_t/T)$. This concept is very similar to Hawking's' "claim that black and white holes are the same. It is the only natural way to make quantum gravity CPT invariant."

## Black Hole Structure using Strings

The quantized string required to reproduce the Standard Model of particle physics has been presented by the author in earlier work". The tripartite string is crucial to understanding the structure of the 'singularity' or 'kernel' of a black (or white) hole. The 'interior' of the 'kernel' of a black hole must exist as an absolutely dense structure in which particles 'crystallize' at the Planck scale without interstitial space. The usual string is an approximation' which assumes an essentially circular shape, which certainly cannot be compacted into an absolutely dense structure. The tripartite string compacts naturally into a tetrahedral structure which would form the kernel of black holes. Note *that such a structure is bilaterally assymmetric, which provides an important foundation through which observations of the Galactic Annihilation Fountain and similar asymmetric ejecta from black holes[5] may be explained. It also easily explains why the entropy of a black hole is related to ¼ its event horizon surface area.*

A string *(or 1-brane)* at the 'kernel' would be 'frozen' in time by the intense forces acting on it. A particle 'frozen' in time *thus drifts away from the 'kernel' and appears to be ejected when it interacts with normal matter* it is locked-in temporally, only to be unlocked in a dual universe at its dual point in time. This observation (admittedly difficult to formally prove!) leads naturally to the notion of ***root***-Mass-present, $M_p$, which accounts for all matter in the universe which is not locked in the interior of black or white holes. The universe is inherently finite infinite mass, infinite spatial dimensions and pointlike objects are all mathematical abstractions which are unphysical. The maximum value of $M_p$ is used to normalize the mass of the universe at any given time, giving $M_{p,n}$ which has a maximum value of 1.

## Universal Mass vs Time

A diagram of the universal Mass vs Time can be constructed from a few basic observations. The total mass contained in white holes is $M_{wh}$ and the total mass accreted onto black holes is $M_{bh}$, where $M_{univ}$ is the total (finite) mass of the universe; $M_p = M_{univ} - M_{wh} - M_{bh}$. Inflationary cosmological models[3] discuss times as short as $10^{-50}$ sec, which is close to the time required for light to propagate across the distance a', which is $10^{-42}$ sec. Quantum gravity will depend on such a time quantum, $T_q$.

The quantization of time is important in cosmology as well as particle physics - in one tick of the quantum clock at the 'instant' of the Big Bang, most white mass was released and inflation began. [The tripartite string[4] requires a quantized spin (or time) to model higher generations of particles.] Some black holes existed from the beginning, as has been established by Hawking, being quite small but gathering mass and *evolving along with their surrounding normal matter*. The Big Bang continued to release matter into the universe at an exponentially-decreasing rate, decomposing into a large number of "shards" which evaporate (or explode) much later. We observe these as supernovae (if at all!), *the existance of which would certainly be crucial to any explanation of "the current epoch of mini-inflation"[6]*. Small, dispersed white-hole evaporation/explosions will continue *to add energy to **our** current space-time* until the final phase in which black holes are the dominant objects in the universe. The final phase of a galaxy being accreted into a *super-massive* black hole proceeds with the rate of accretion increasing as the event horizon grows - until there is no more galactic matter in the vicinity. On a universal logarithmic time scale, this would appear as the relation $M_{univ}-M_{bh}$ curving downward.

Define $T_n = K*\ln(T/T_o)$ and take $T_0 = 10^{-6}$ second, the estimated time of chiral symmetry breaking[7,3]. With $T_{begin} = \ln(T_q/1)/42$ and $T_{present} \cong 16$ billion years gives $T_{end} \cong 200$ trillion years.

## Conclusions

The *cosmological constant and* 'flat' universe is explained as a direct consequence of an equation of the form: $M^2_{p,n} + T^2_n \approx \Omega_m + \Omega_\Lambda = 1$, *which indicates that the square root of mass is an important dimension, meaning that mass is essentially area-like and related to I-brane **area**[8], orthonormally to the string's tension, the basis for energy. Mass is likely also an exponential function of area as in ref 9. No critical mass is required by this model. The observed "current epoch of mini-inflation"[6] and asymmetric black-hole ejecta[5] are consistent with this model.*